\documentclass[twocolumn,prl,aps,superbib,tightenlines,floatfix]{revtex4}
\usepackage{amsfonts}
\usepackage{amsmath}
\usepackage{amssymb}
\usepackage{graphicx}
\begin{document}
\bibliographystyle{apsrev}
%\draft
%\wideabs
\title{Inelastic current-voltage characteristics of atomic and molecular junctions}
\author{Yu-Chang Chen}
\affiliation{Department of Physics, University of California, San Diego, La Jolla, CA 92093-0319}
\author{Michael Zwolak}
\affiliation{Department of Physics, California Institue of Technology, Pasadena, California 91125}
\author{Massimiliano Di Ventra\cite{MD}}
\affiliation{Department of Physics, University of California, San Diego, La Jolla, CA 92093-0319 }

\begin{abstract}
We report first-principles calculations of the inelastic current-voltage (I-V) characteristics of a gold
point contact and a molecular junction in the nonresonant regime. Discontinuities in the I-V curves appear in correspondence to the normal
modes of the structures. Due to the
quasi-one-dimensional nature of these systems, specific modes with large longitudinal component dominate the inelastic I-V curves. In the case of the gold point contact, our results are in good agreement with recent experimental data.
For the molecular junction, we find that the inelastic I-V curves are quite sensitive to the
structure of the contact between the molecule and the electrodes thus providing a powerful tool to extract the
bonding geometry in molecular wires.
\end{abstract}
\maketitle
%]\bigskip
%\narrowtext

Inelastic scattering between electrons and phonons in a current-carrying
wire is a source of energy dissipation for electrons. However, it can also
yield a lot of information on the underlying atomic structure of the wire.
This information can be extracted indirectly from the discontinuities in
conductance that occur when the external bias is large enough to excite
discrete vibrational modes of the wire.~\cite{lorente} Recent experiments on
transport properties of atomic~\cite{agrait} and molecular~\cite{park,smit}
junctions have indeed revealed such inelastic features. It is, however, not
straightforward to relate these features to specific vibrational modes. A
nanoscale junction (often described as a quasi-one-dimensional system) with $%
N$ atoms supports $3N$ vibrational modes. In a strictly one-dimensional
system, only longitudinal modes can be excited via electronic coupling.
However, the modes of a realistic junction are not necessarily purely
transverse or purely longitudinal with respect to the direction of current
flow.~\cite{chen1,todorov03,troisi} Therefore, the inelastic current-voltage
(I-V) characteristics are likely to depend strongly on the detailed atomic
structure of the full system. This is particularly relevant for molecular
junctions for which the contact geometry between the molecule and the bulk
electrodes is difficult to control in experiments.~\cite{tao,diventra1}

In this letter we first derive an expression for the inelastic current in a
current-carrying system in terms of scattering wavefunctions. This
expression allows us to study the inelastic I-V characteristics of a given
nanoscale junction using first-principles approaches. As an example we study
the effect of vibrations on the electron dynamics in a gold point contact
and a single-molecule junction. For the gold point contact, the magnitude of
the calculated inelastic current as well as its onset compare very well with
recent experimental results.~\cite{agrait} For the molecular junction, we
analyze the case in which the molecule is equally bonded to the two bulk
electrodes and the case in which the two contacts are different. We find
that the inelastic I-V characteristics are very different in the two cases.
This result shows that inelastic spectroscopy could be used quite
effectively to extract information on the contact geometry of molecular
wires.

Let us start by deriving an expression for the inelastic current. We assume
that the phonon distribution is at equilibrium at all (small) biases, thus
neglecting local heating.~\cite{chen1} We have previously shown that for
small biases this effect is small, provided good thermal contacts exist
between the nanostructures and the bulk electrodes.~\cite{chen1} The
many-body Hamiltonian of the system is (atomic units are used throughout
this paper)~\cite{chen1}
\begin{equation}
H=H_{el}+H_{vib}+H_{el-vib},  \label{ham}
\end{equation}
where $H_{el}$ is the electronic part of the Hamiltonian; $H_{vib}=\frac{1}{2%
}\sum\limits_{i,\mu\in vib}\dot{q}_{i\mu}^{2}+\frac{1}{2}\sum\limits_{i,\mu%
\in vib}\omega_{i\mu}^{2}q_{i\mu}^{2}$ is the ionic contribution where $%
q_{i\mu}$ is the normal coordinate and $\omega_{i\mu}$ is the normal
frequency corresponding to the $i$-th ion and $\mu$-th component; finally, $%
H_{el-vib}$ describes the electron-ion interaction and has the following
form:

\begin{align}
H_{el-vib} & =\sum_{\alpha,\beta}\sum_{E_{1},E_{2}}\sum_{i\mu,j\nu\in vib}
\notag  \label{vib} \\
& \sqrt{\frac{\hbar}{2\omega_{j\nu}}}A_{i\mu,j\nu}J_{E_{1},E_{2}}^{i\mu,%
\alpha\beta}a_{E_{1}}^{\alpha\dag}a_{E_{2}}^{\beta}\left( b_{j\nu
}+b_{j\nu}^{\dag}\right) ,
\end{align}
where $\alpha=L,R$; $a_{E}^{\alpha}$ and $b_{j\nu}$ are the electron and
phonon annihilation operators, respectively, satisfying the usual
commutation relations. $A_{i\mu,j\nu}$ are the matrix elements of the
transformation from cartesian coordinates to normal coordinates, and $%
J_{E_{1},E_{2}}^{i\mu,\alpha\beta}$ is the electron-phonon coupling constant
which can be directly calculated from the scattering wave-functions~\cite%
{chen1}
\begin{equation}
J_{E_{1},E_{2}}^{i\mu ,\alpha \beta }=\int d\mathbf{r}\int d\mathbf{K}%
_{\Vert }\Psi _{E_{1}}^{\alpha \ast } \left( \mathbf{r},\mathbf{K}_{\Vert
}\right) \partial _{\mu}V^{ps}\left( \mathbf{r},\mathbf{R}_{i}\right) \Psi
_{E_{2}}^{\beta }\left( \mathbf{r},\mathbf{K}_{\Vert }\right) ,
\label{couplej}
\end{equation}%
where we have chosen to describe the electron-ion interaction with
pseudopotentials $V^{ps}\left( \mathbf{r},\mathbf{R}_{i}\right) $ for each
i-th ion.~\cite{diventra}

Similar to what has been done in Ref.~\onlinecite{todorov03}, we treat the
electron-phonon interaction to first-order perturbation theory.~\cite{prec00}
Due to the orthogonality condition between phonon states, higher harmonics
for each phonon mode appear only in third-order perturbation theory and are
therefore small.~\cite{vivek} We develop the full many-body wavefunctions in
terms of the states $\left\langle \Psi_{E}^{L(R)};n_{j\nu }\right\vert
=\left\langle \Psi_{E}^{L(R)}\right\vert \otimes\left\langle
n_{j\nu}\right\vert $. The single-particle electronic state is described by $%
\Psi_{E}^{L(R)}\left( \mathbf{r,\mathbf{K}_{\Vert}}\right)$, corresponding
to electrons incident from the left (right) electrodes with energy E and
momentum $\mathbf{K}_{\Vert}$ parallel to the electrode surface.~\cite%
{diventra} These electronic states are calculated self-consistently by means
of a scattering approach within the density functional theory of
many-electron systems.~\cite{diventra} The phonon state is described by $%
\left\langle n_{j\nu}\right\vert $, where $n_{j\nu}$ is the number of
phonons in the $j\nu$-th normal mode.

\begin{align}
\left\vert \delta \Psi _{E}^{\alpha };n_{j\nu }\right\rangle &
=\lim_{\epsilon \rightarrow 0^{+}}\sum_{\alpha ^{\prime
}=L,R}\sum_{j^{\prime }\nu ^{\prime }}\int dE^{^{\prime }}D_{E^{^{\prime
}}}^{\alpha ^{\prime }}  \notag \\
& \frac{\left\langle \Psi _{E^{\prime }}^{\alpha ^{\prime }};n_{j^{\prime
}\nu ^{\prime }}\left\vert H_{el-vib}\right\vert \Psi _{E}^{\alpha };n_{j\nu
}\right\rangle \left\vert \Psi _{E^{\prime }}^{\alpha ^{\prime
}};n_{j^{\prime }\nu ^{\prime }}\right\rangle }{\varepsilon (E,n_{j\nu
})-\varepsilon (E^{\prime },n_{j^{\prime }\nu ^{\prime }})-i\epsilon }.
\label{perturbation}
\end{align}
In the above expression $D_{E}^{\alpha}$ is the partial density of states
corresponding to the current-carrying states $\Psi_{E}^{\alpha}$ and $%
\varepsilon(E,n_{j\nu})=E+\left( n_{j\nu }+1/2\right) \hbar\omega_{j\nu}$ is
the energy of state $\left\vert \Psi _{E}^{\alpha};n_{j\nu}\right\rangle $.
We have also assumed that the electrons rapidly thermalize into the bulk
electrodes so that their statistics are given by the equilibrium Fermi-Dirac
distribution, $f_{E}^{L(R)}=1/(\exp[(E-E_{FL(R)})/k_{B}T_{e}]+1)$ with
chemical potential $E_{FL(R)}$ deep into the left (right) electrode.~\cite%
{chen2} Using $\lim_{\epsilon\rightarrow 0}\frac{1}{z-i\epsilon}=P(\frac{1}{z%
})+i\pi\delta(z)$, the first-order correction $\left\vert
\delta\Psi_{E}^{\alpha};n_{j\nu}\right\rangle $ assumes the following form:

\begin{gather}
\left\vert \delta \Psi _{E}^{R};n_{j\nu }\right\rangle =i\pi
\sum_{i\mu \in
vib}\sqrt{\frac{\hbar }{2\omega _{j\nu }}}A_{i\mu ,j\nu }  \notag \\
\lbrack D_{E+\hbar \omega _{j\nu }}^{L}\sqrt{\left\langle n_{j\nu
}\right\rangle f_{E}^{R}(1-f_{E+\hbar \omega _{j\nu }}^{L})}\cdot   \notag \\
J_{E+\hbar \omega _{j\nu },E}^{i\mu ,LR}\left\vert \Psi _{E+\hbar
\omega
_{j\nu }}^{L};n_{j\nu }-1\right\rangle +  \notag \\
D_{E-\hbar \omega _{j\nu }}^{L}\sqrt{\left( 1+\left\langle n_{j\nu
}\right\rangle \right) f_{E}^{R}(1-f_{E-\hbar \omega _{j\nu
}}^{L})}\cdot
\notag \\
J_{E-\hbar \omega _{j\nu },E}^{i\mu ,LR}\left\vert \Psi _{E-\hbar
\omega _{j\nu }}^{L};n_{j\nu }+1\right\rangle ],  \label{wavefn}
\end{gather}%

where $\left\langle n_{j\nu }\right\rangle =1/\left[ \exp \left( \hslash
\omega _{j\nu }/k_{B}T_{w}\right) -1\right] $ is the Bose-Einstein
distribution per mode at a given wire temperature $T_{w}$, and $\left\langle
{}\right\rangle $ indicates the statistical average. The above expression
allows us to calculate the inelastic current. It is evident from Eq.~\ref%
{wavefn} that for a fixed partial density of states, the magnitude of the
inelastic current is determined by the coupling constant $%
J_{E_{1},E_{2}}^{i\mu ,\alpha \beta }$ and the transformation matrix $%
\mathbf{A}=\left\{ A_{i\mu ,j\nu }\right\} $ which contains the information
on the geometry of the structure and hence on the character (transverse
versus longitudinal) of the different modes.~\cite{longvstrans} We will be
concerned with the \textit{extra} inelastic current due to the vibrational
modes of the atoms of the nanoscale constriction with respect to the
continuum spectrum of modes of the bulk electrodes.~\cite{prec2} If the
electronic temperature $T_{e}$ is zero, then, for an external bias $V$, only
those normal modes with eigenenergies $\hbar \omega _{j\nu }<eV$ can be
excited and contribute to Eq.~\ref{wavefn}. In addition, due to our
assumption of negligible local heating, the averaged number of phonons $%
\left\langle n_{j\nu } \right\rangle $ is zero for all normal
modes. In this case the first-order correction to the current induced by
electron-phonon interaction assumes the following simple form:

\begin{align}
\delta I & =-i\int_{E_{FL}}^{E_{FR}}dE\int d\mathbf{R}\int d\mathbf{K}%
_{\Vert}  \notag \\
& [\left( \delta\Psi_{E}^{R}\right)
^{\ast}\partial_{z}\delta\Psi_{E}^{R}-\partial_{z}\left(
\delta\Psi_{E}^{R}\right) ^{\ast}\delta\Psi_{E}^{R}\mathbf{)]},
\label{current}
\end{align}
where only the left-travelling electronic states contribute (if the left
electrode is positively biased).

\begin{figure}[tbp]
\includegraphics[width=.48\textwidth]{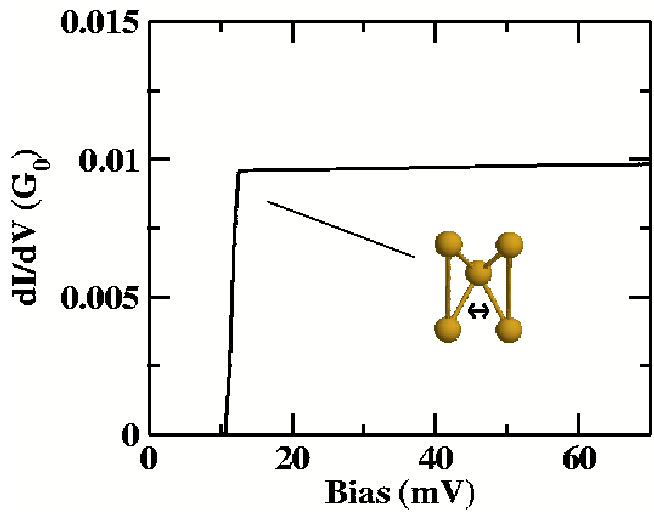}
\caption{Absolute value of the differential conductance due to
electron-phonon interaction as a function of bias for a gold point contact.
Two normal modes corresponding to transverse vibration of the gold atom
between the electrodes have energies of 10.8 meV, which are close in energy
to the longitudinal mode at 11.5 meV (shown in the figure).}
\label{fig1}
\end{figure}

We are now ready to use the above expression to study inelastic scattering
in specific systems. We choose to first study a gold point contact for which
experimental results are available~\cite{agrait} and then discuss the case
of a molecular junction. In Fig.~\ref{fig1} we plot the inelastic
conductance for a single gold atom.

In the absence of inelastic scattering and for the bias range of Fig.~\ref%
{fig1}, the I-V characteristics of this system are linear with differential
conductance $G\simeq1.1$ $G_{0}$, where $G_{0}=2e^2/h$.~\cite{chen1} When
electron-phonon interactions are considered, two transverse modes with
energy $\hbar\omega\simeq10.8$ meV are first excited with increasing bias.
However, due to their transverse character, these modes contribute
negligibly to the inelastic current as determined by the product between the
transformation matrix $\mathbf{A}$ and the coupling constant $%
J_{E_{1},E_{2}}^{i\mu,\alpha\beta}$. An abrupt change in differential
conductance appears at $V\simeq11.5$ $mV$ corresponding to the excitation of
a longitudinal vibrational mode (see schematic in Fig.~\ref{fig1}).~\cite%
{prec} Both the onset bias as well as the change in conductance (of about
1\%) are in good agreement with experimental reports on gold point contacts.~%
\cite{agrait} The longitudinal and transverse modes are very close in energy
but mainly the longitudinal mode contributes to the inelastic current,~\cite%
{todorov03,troisi} so that in experiments the transverse ones would not be
easily resolved. This is even more evident in the case of the molecular
junction.

\begin{figure}[tbp]
\includegraphics[width=.48\textwidth]{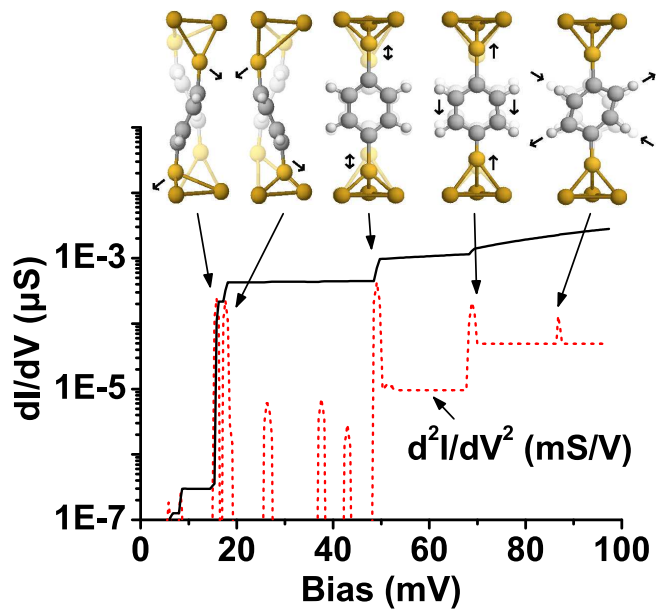}
\caption{Absolute value of the differential conductance due to
electron-phonon interaction as a function of bias for a symmetric molecular
junction. The derivative of the conductance with respect to bias is also
shown (a broadening of 1 meV is introduced). The schematics show only the
modes that contribute the most to the inelastic current.}
\label{fig2}
\end{figure}

In Fig.~\ref{fig2} we plot the inelastic conductance in the case in which a
phenyldithiolate molecule forms symmetric contacts on both sides of the
junction, i.e., each S atom is bonded to a flat surface. In this case there
is a total of 14 modes with energy less than 100 meV. A prominent change in
conductance occurs at a bias of about 18 mV, i.e., at a bias large enough to
excite two modes with large longitudinal component (see Fig.~\ref{fig2}).~%
\cite{longvstrans1} The inelastic contribution from two transverse modes at
lower bias~\cite{chen1} is almost four orders of magnitude smaller.
Similarly, three quasi-transverse modes with energies between 20 and 50 meV
contribute negligibly to the inelastic conductance. They only appear as
small features in the second derivative of the current with respect to the
bias (see Fig.~\ref{fig2}).~\cite{prec0} It is likely that due to noise and
other effects, such modes would not be resolved in experiments. Increasing
the bias further, a second large step in the absolute value of the
conductance is found at about 50 mV (see Fig.~\ref{fig2}). This again
corresponds to a predominantly longitudinal mode (see schematic in Fig.~\ref%
{fig2}). This mode is then followed by others that have both a transverse
and a longitudinal component. The magnitude of the conductance steps depends
on the relative amount of the two components as well as the product between
the transformation matrix $\mathbf{A}$ and the coupling constant.~\cite%
{longvstrans1}

\begin{figure}[tbp]
\includegraphics[width=.48\textwidth]{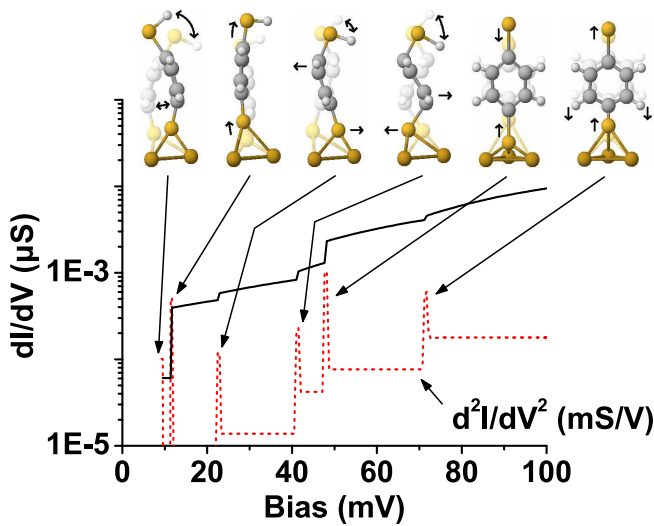}
\caption{Absolute value of the differential conductance due to
electron-phonon interaction as a function of bias for a molecular junction
with asymmetric contacts. The derivative of the conductance with respect to
bias is also shown (a broadening of 1 meV is introduced). The schematics
show only the modes that contribute the most to the inelastic current.}
\label{fig3}
\end{figure}

We conclude by showing how sensitive the inelastic current is to any change
in the bonding properties of the molecule to the electrodes. We illustrate
this in Fig.~\ref{fig3} where we plot the inelastic conductance for the same
molecule but with one of the S atoms bonded to a H atom which, in turn, is
not bonded to the nearby surface (see schematics in Fig.~\ref{fig3}).~\cite%
{prec3} Such a configuration could be easily realized in experiments.~\cite%
{tao} In the present case there are 13 modes below 100 meV, of which only
six have large longitudinal component (shown in Fig.~\ref{fig3})~\cite%
{longvstrans1} and contribute to small steps in the inelastic conductance,
with the one at about 11 meV showing the largest relative contribution. All
other modes contribute negligibly to the current. Comparing Figs.~\ref{fig2}
and~\ref{fig3} it is clear that contact geometry affects considerably the
I-V characteristics of the system both in the position of the inelastic
discontinuities as well as in their relative magnitude. This fact, however,
can be used advantageously to extract \textit{a posteriori} the contact
structure of molecular junctions thus providing a powerful diagnostic tool
for nanoscale electronics.

We acknowledge support from the NSF Grant Nos. DMR-01-02277 and
DMR-01-33075, and Carilion Biomedical Institute. Acknowledgment is also made
to the Donors of The Petroleum Research Fund, administered by the American
Chemical Society, for partial support of this research. One of us (MZ)
acknowledges partial support from an NSF Graduate Fellowship.

\begin{references}

\bibitem[*]{MD}E-mail address: diventra@physics.ucsd.edu

\bibitem{lorente} N. Lorente, M. Persson, L.J. Lauhon, W. Ho, Phys. Rev. Lett. {\bf 86}, 2593 (2001).

\bibitem {agrait}N. Agra\"{\i}t, C. Untiedt, G. Rubio-Bollinger, and S. Vieira,
Phys. Rev. Lett. \textbf{88}, 216803 (2002).

\bibitem {park}H. Park et al. Nature, \textbf{407},57 (2000); L.H. Yu and D. Natelson, Nano Letters, {\bf 4},
79 (2004).

\bibitem {smit}R.H.M. Smit, Y. Noat, C. Untiedt, N.D. Lang, M.C. van Hemert, and J.M. van Ruitenbeek,
Nature \textbf{419}, 906 (2002).

\bibitem {chen1}Y.-C. Chen, M. Zwolak, and M. Di Ventra, Nano Lett. {\bf 3}, 1691 (2003).

\bibitem {todorov03}M.J. Montgomery, J. Hoekstra, T.N. Todorov, and A. Sutton, J. Phys.:
Cond. Mat. \textbf{15}, 731 (2003).

\bibitem{troisi} A. Troisi, M.A. Ratner, and A. Nitzan, J. Chem. Phys. {\bf 118}, 6072 (2003).

\bibitem{tao} X. Xiao, B. Xu, and N.J. Tao, Nano Lett. {\bf 4}, 267 (2004).

\bibitem{diventra1} M. Di Ventra, S.T. Pantelides, and N.D. Lang, Phys. Rev. Lett. \textbf{84}, 979 (2000).

\bibitem {diventra}N.D. Lang, Phys. Rev. B \textbf{52}, 5335 (1995); M. Di
Ventra and N.D. Lang, Phys. Rev. B \textbf{65}, 045402 (2002); Z. Yang, A.
Tackett, and M. Di Ventra, Phys. Rev B. \textbf{66}, 041405 (2002).

\bibitem{prec00} Since we are in the regime of weak electron-phonon coupling we neglect the elastic phonon renormalization
of the electronic spectrum (see e.g. G.D. Mahan, {\it Many Particle Physics}, 2nd Edition (Plenum Publishers, NY,
1990), p. 285).

\bibitem {vivek}V. Aji, J.E. Moore, and C.M. Varma, cond-mat/0302222;
A. Mitra, I. Aleiner, A.J. Millis, cond-mat/0311503.

\bibitem {chen2}Y.-C. Chen and M. Di Ventra, Phys. Rev. B \textbf{67},153304 (2003).

\bibitem{longvstrans} A quantitative way of distinguishing between longitudinal versus transverse character of
each mode is to sum up the absolute values of all the $z$-component (i.e., along the current flow) matrix elements
of $\mathbf{A}$.

\bibitem{prec2} We have employed Hartree-Fock total energy calculations
[see, e.g.,  J. A. Boatz and M. S. Gordon, J.Phys. Chem., {\bf 93}, 1819 (1989)]
to evaluate the vibrational modes of the
single-molecule junctions and the gold point contact. For these calculations,
the assumed surface orientation is [111] represented by a triangular pad of gold atoms with infinite mass
(see insets of Figs.~\ref{fig1} and~\ref{fig2}). The structures were relaxed starting from an  initial geometry where the single gold atom of the point contact faces the center of the triangular pad at a distance of 2.3\AA; and for the single-molecule junction with symmetric
contacts the initial S-surface distance is 2.4\AA. In the case of asymmetric contacts the S atom on one side of the
junction is bonded to a H atom (see insets of Fig.~\ref{fig3}). This thiol termination is assumed to be far from the nearby surface so that the modes of this structure are not influenced by the gold atoms of the surface.

\bibitem{prec} Note that the inelastic correction to the current is actually negative. For convenience
we plot in this paper the absolute value of this correction.

\bibitem{longvstrans1} In the molecular junction cases, the dominant net contribution to the electron-phonon
coupling is from the S atoms and the adjacent C atoms (i.e., the two C atoms that form a straight line with the S atoms). This implies that
those modes that do not have large
displacements of these atoms along the $z$ direction will contribute negligibly to the inelastic current, even if
the global character of the mode is longitudinal as inferred from the procedure outlined in
Ref.~\onlinecite{longvstrans}. This is, e.g. the case for the mode at about 90 meV shown schematically in
Fig.~\ref{fig2}.

\bibitem{prec0} A broadening of 1 meV is introduced in all calculations to make the second derivative of the
inelastic current as a function of bias finite at the conduction jumps.

\bibitem{prec3} In order to facilitate the comparison with the previous case, we have assumed in this calculation the same coupling constants,
partial density of states and unperturbed current as for the symmetric junction case.

\end{references}
\end{document}